\documentstyle[editedvolume]{crckapb}

\begin{opening}
\title{Cygnus X-1: X-ray Emission Mechanism and Geometry}

\author{Wei Cui}
\institute{MIT, Cambridge, MA 02139, USA}
\author{S.~N.~Zhang}
\institute{NASA/MSFC, Huntsville, AL 35812, USA}
\author{J.~H.~Swank}
\author{X.-M.~Hua}
\author{K.~Ebisawa}
\institute{NASA/GSFC, Greenbelt, MD 20771, USA}
\author{T.~Dotani}
\institute{ISAS, Yoshinodai, Sagamihara, Kanagawa, 229, Japan}

\end{opening}

\runningtitle{X-ray Emission from Cyg X-1}

\begin{document}


\section{Introduction}
\noindent
On 1996 May 10, the All-Sky Monitor aboard RXTE revealed that Cyg~X--1 started
a transition from its hard state to soft state (Cui 1996). Throughout this 
interesting episode, snapshots were taken with more sensitive detectors on 
ASCA, RXTE, and CGRO to monitor the temporal and spectral variability of the 
source over a broad energy range.

The results (both spectral and temporal) from these observations seem to 
converge toward a self-consistent picture, which not only specifies the X-ray 
emission geometry but also qualitatively describes how the geometry evolves 
during the transition.

\section{Results and Discussion}
\noindent
The X--ray spectrum of Cyg~X--1 extends beyond 100 keV in both states. The
spectral shape is rather complicated in details. At low energies, it is 
dominated by an ultra-soft component, which is thought to be the emission 
from an optically thick, geometrically thin disk. During the transition, the 
color temperature varies by a factor of 2 -- 3, and the luminosity (of this 
component) by a factor of 3, indicating that the inner disk edge moves about 
3 times closer to the black hole as the soft state is approached (Zhang et al.
1997). At high energies the
spectrum can be described by a Comptonized spectrum. It is steeper in the
soft state, implying a smaller Comptonizing corona, perhaps due to more 
efficient cooling (Cui et al. 1997a). Such interpretation is strongly 
supported by the measured hard X-ray lags: much smaller in the soft state,
as well as observed shape of power-density spectra (Cui et al. 1997a,c).
The measured coherence function between various energy bands is nearly 
unity in both states (Vaughan \& Nowak 1997; Cui et al. 1997c), but is less
during the transition (Cui et al. 1997c), implying that the corona indeed 
varies during such a period.

A Compton reflection ``bump'' is apparent on the spectrum, as well as the 
presence of an iron emission line. In the soft state, the reflecting medium
is highly ionized and the solid angle subtended is smaller than in the hard
state (Ebisawa et al. 1996; Cui et al. 1997b). This is consistent with the
reflection occuring mostly at the inner disk edge that is closer to the black 
hole (thus hotter) in the soft state. The iron line centers at $\sim$6.4 keV 
in the hard state, probably due to neutral irons (Ebisawa et al. 1996), while 
in the soft state it is mostly due to He-like irons (Cui et al. 1997b). This 
suggests that the line emission also originates in the innermost region of 
the disk. The broader line profile in the soft state supports this scenario.

Attempts have been made, for the first time, to simultaneously model the
observed spectral and timing properties of Cyg X-1 (Hua et al. 1997). 
The model invokes an extended hot Comptonizing
corona with a non-uniform density distribution. It reproduces the observed
spectral shape well, and more importantly it is capable of
explaining the frequency-dependence of measured hard lags, which has been
a ``dilema'' for Compton models over the past decade (Miyamoto et al. 1988). 
It now becomes clear that
such dependence is sensitive to the density distribution of the corona. 
Therefore, phase lag measurement may ultimately provide insights into the 
dynamics of hot corona. In the future, more physics need to be incorporated in 
the model to account for such features as the reflection bump and iron line.

\section{References}
Cui,~W. 1996, IAUC. 6404 \\
Cui,~W., et al. 1997a, ApJ, 474, L57 \\
Cui,~W., Ebisawa,~K., Dotani,~T. \& Kubota,~A. 1997b, ApJ, submitted \\
Cui,~W., Zhang,~S.~N.,Focke,~W., \& Swank,~J. 1997c, ApJ, 484, 383 \\
Ebisawa,~K. et al. 1996, ApJ, 467, 419 \\
Hua,~X.-M., Kazanas,~D., \& Cui,~W. 1997, ApJ, submitted \\
Miyamoto,~S., et al. 1988, Nature, 336, 450 \\
Vaughan,~B.~A., \& Nowak,~M.~A. 1997, ApJ, 474, L43 \\
Zhang,~S.~N., et al. 1997, ApJ, 477, L95
\end{document}